\documentclass[apjl]{myemulateapj}
 \usepackage{apjfonts}
\usepackage{graphicx}

\begin{document}
\title{The first determination of the viscosity parameter in the circumstellar disk of a Be Star}
\author{
Alex C. Carciofi\altaffilmark{1},
Jon E. Bjorkman\altaffilmark{1,2},
Sebasti\'an A. Otero\altaffilmark{3},
Atsuo T. Okazaki\altaffilmark{4},
Stanislav~\v{S}tefl\altaffilmark{5},
Thomas Rivinius\altaffilmark{5},
Dietrich Baade\altaffilmark{6},
and
Xavier Haubois\altaffilmark{1}
}
\altaffiltext{1}{Instituto de Astronomia, Geof\'isica e Ci\^encias Atmosf\'ericas, Universidade de S\~ao Paulo, Rua do Mat\~ao 1226, Cidade Universit\'aria, 05508-900, S\~ao Paulo, SP, BRAZIL, carciofi@usp.br}
\altaffiltext{2}{Ritter Observatory, Department of Physics \& Astronomy, Mail Stop 113, University of Toledo, Toledo, OH 43606, jon@physics.utoledo.edu}
\altaffiltext{3}{Asociaci\'on Cielo Sur Grupo Wezen 1 88, Buenos Aires, Argentina}
\altaffiltext{4}{Faculty of Engineering, Hokkai-Gakuen University, Toyohira-ku, Sapporo 062-8605, Japan
}
\altaffiltext{5}{European Organisation for Astronomical Research in the Southern Hemisphere, Casilla 19001, Santiago 19, Chile}
\altaffiltext{6}{European Organisation for Astronomical Research in the Southern Hemisphere, Karl-Schwarzschild-Str.~2, 85748 Garching bei M\"unchen, Germany 
}

\begin{abstract}
Be stars possess gaseous circumstellar decretion disks, which are well described using standard $\alpha$-disk theory.
The Be star 28\,CMa recently underwent a long outburst followed by a long period of quiescence, during which the disk dissipated. Here we present the first time-dependent models of the dissipation of a viscous decretion disk.
By modeling the rate of decline of the $V$-band excess, we determine that the viscosity parameter $\alpha=1.0\pm0.2$, corresponding to a mass injection rate $\dot{M}=(3.5\pm 1.3) \times 10^{-8}\ M_\sun\,\mathrm{yr}^{-1}$.
Such a large value of $\alpha$ suggests that the origin of the turbulent viscosity is an instability in the disk whose growth is limited by shock dissipation.  The mass injection rate is more than an order of magnitude larger than the wind mass loss rate inferred from UV observations, implying that the mass injection mechanism most likely is not the stellar wind, but some other mechanism.
\end{abstract}

\keywords{circumstellar matter --- stars: individual (28\,CMa) --- stars: activity --- stars: emission-line, Be}

\section{Introduction} \label{intro}

It is now generally accepted that Classical Be stars possess gaseous circumstellar disks that are responsible for their excess emission \citep[for reviews, see][]{por03,baa01}.  Furthermore spectroastrometry and optical interferometic measurements indicate that this material is in Keplerian rotation \citep{oud11,kra11}.  Consequently, as detailed in the review of \citet{car11}, the steady-state viscous decretion disk model \citep{lee91} is capable of explaining many of the (time-averaged) observational properties of Be star disks.  

Most likely these disks are fed by mass-loss from their rapidly rotating central stars.  Both observational \citep{riv98,nei02} and theoretical \citep{and86,cra09} studies suggest that non-radial pulsations may play an important role.  However, to create a viscous decretion disk, this material must be placed into orbit.  Since the stars most likely are rotating at less than their critical speed \citep[e.g.,][]{cra05}, the mass loss mechanism must add some angular momentum before injecting the material into to the disk.  From a theoretical point of view, this requirement is problematic, so, as yet, no entirely satisfactory mechanism has been proposed \citep[see][for a detailed assessment]{owo06}.

Regardless of how material is injected into the disk, if this material is in orbit, turbulent viscosity will redistribute it on a viscous diffusion timescale, producing an outflowing decretion disk, as long as the disk is fed continually.  However, Be stars exhibit a wide range of variability, ranging from emission line profile variations \citep[e.g, $V/R$ variability, indicative of density waves within the disk; ][]{ste09,car09}, to photometric and polarimetric variability, indicating that frequently the disks are not fed at a constant rate.  In some cases, the disk can disappear for decades before being rebuilt \citep[see review by][]{und82,wis10}.  Oftentimes, this rebuilding phase occurs in a series of outbursts, during which the disk apparently is fed for a short period of time followed by a period of quiescence, during which the nascent disk can partially reaccrete onto the star.  Similar outbursts can occur while the disk is disappearing.

Once such example is the Be star 28\,CMa.  Over the past 40 years, 28\,CMa has exhibited quasi-regular outbursts, every 8 years or so, when the star brightens by about half a magnitude in the $V$-band.  This usually occurs via multiple individual injections (e.g., the rapid rises seen in Fig. 1), during which the star brightens at a rate of $0.02\pm0.01\,\mathrm{mag\,d}^{-1}$, due to free-bound and free-free radiation from the disk.  The outbursts last for about two years \citep[see][for a detailed discussion of the long-term photometric  variations of 28\,CMa]{ste03}.  A particularly well-monitored outburst occurred from 2001 to 2004, and this outburst was followed by a long quiescent phase from 2004 to 2009.  Furthermore it appears that during this quiescent phase, very little additional material was injected into the disk (only a few minor events).  Thus nature has provided us the perfect experiment to study how Be star disks dissipate.

If we accept as a premise that the viscous decretion disk model is correct (and that no other mechanisms participate in the disk mass budget), then the timescale of the post-outburst photometric decline is controlled solely by the disk viscosity.  In this paper, we model the $V$-band light curve with the goals of: 1) directly measuring the disk viscosity parameter, $\alpha$, and 2) further testing the viscous decretion disk model by comparing its time-dependent predictions of the disk dissipation against the detailed shape of the photometric light curve.

\begin{figure*}
\centerline{\includegraphics[width=0.85\linewidth]{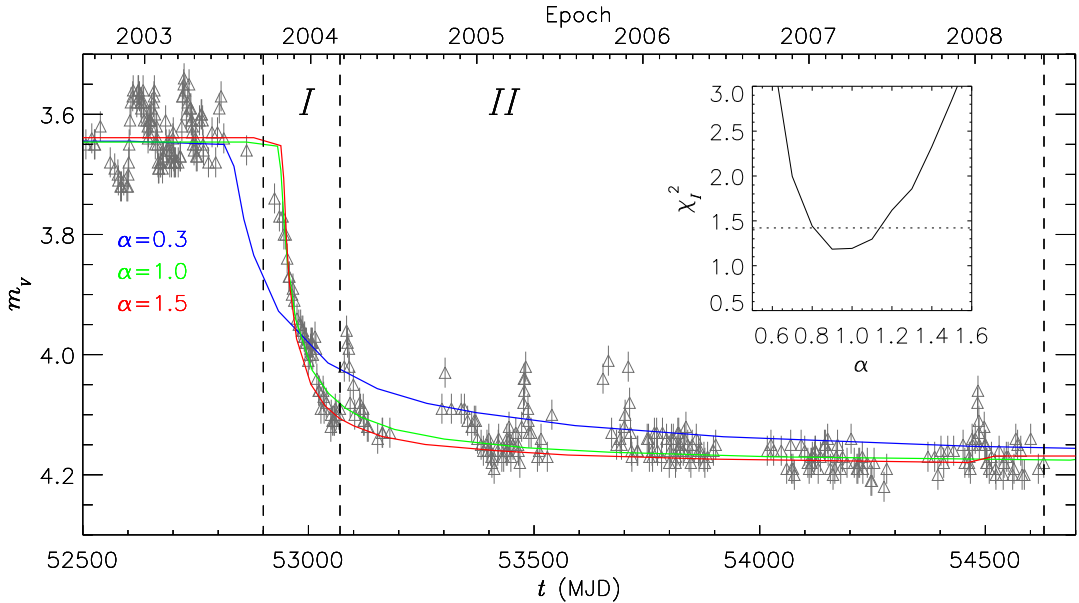}}
\caption{$V$-band Light Curve.  Visual observations of 28\,CMa (grey triangles) are shown in comparison to our model fits for different values of $\alpha$, as indicated.  Phase~I (MJD = 52900--53070) is the initial decline, which was used to determine the value of $\alpha$. Phase~II (MJD = 53070--54670) is the slow disk-draining phase.
The inset shows the reduced chi-squared of the Phase~I fit for different values of $\alpha$. The horizontal dotted line indicates the 90\% confidence level.
}
\label{fig:LightCurve}
\end{figure*}

\section{Observations}
Visual photometric observations of 28\,CMa ($\omega$\,CMa, HD\,56139, HR\,2749; B2--3\,IV--Ve; $m_V = 3.6$ to 4.2, $B-V =-0.1$ to $-0.2$) were made from early 2003 to late 2009 using a modified version of the \citet{arg1843} method in which the estimation of the variable star's brightness is performed by comparison with two non-variable stars, one brighter, the other fainter than the target.  The accuracy depends on the availability of nearby comparison stars with similar brightness and color.  Since 28\,CMa is variable, several pairs of comparison stars are required. 
Based on comparisons with simultaneous photoelectric $V$-band photometry, the error is typically less than 0.05 mag \citep*{ote01,ote06}.  However, based on the typical dispersion of the 28\,CMa observations, we adopt a uniform 1-$\sigma$ error of 0.03 mag.
 The resulting $V$-band light curve for 28\,CMa is shown in Figure~\ref{fig:LightCurve}.

\section{Modeling}
To model the circumstellar disk of 28\,CMa, we require physical parameters for the central star, including the inclination angle of the system.  Fortunately, \citet{mai03} performed an extensive investigation of the non-radial pulsation modes of this star.  Using detailed fits of the line profile variability, they conclude that the star is viewed nearly pole-on at an inclination angle $i = 15^\circ$.  Table~\ref{tab:StellarParameters} lists their recommended values for the stellar parameters.  

\begin{deluxetable}{ll}
\tablecaption{Stellar Parameters \label{tab:StellarParameters}}
\tablewidth{0pt}
\tablehead{
\colhead{Parameter} &
\colhead{Value} \\[-10pt]
}
\startdata
$L$               \dotfill    & $5224\,L_\sun$ \\
$T_\mathrm{pole}$ \dotfill    & $22000\ {\rm K}$ \\
$R_\mathrm{pole}$ \dotfill    & $6.0\ R_\sun$ \\
$\log g_\mathrm{pole}$ \ldots & $3.84$ \\
$M$               \dotfill    & $9\ M_\sun$ \\
$V_\mathrm{rot}$  \dotfill    & $350\ {\rm km\,s}^{-1}$ \\
$V_\mathrm{crit}$ \dotfill    & $436\ {\rm km\,s}^{-1}$ \\
$R_\mathrm{eq}$   \dotfill    & $7.5\ R_\sun$ \\
$i$               \dotfill    & $15^\circ$ \\[-6pt]
\enddata
\end{deluxetable}

Given the stellar parameters, we now model the disk dissipation of 28\,CMa. To do so, we combine a time-dependent calculation of the disk surface density with a NLTE radiative transfer calculation of the emergent flux at selected times.  The disk surface density is modeled with a viscous decretion disk \citep*{lee91} using the time-dependent hydrodynamics code {\sc singlebe}  \citep{oka02,oka07}, which permits time-variable mass injection at a point placed just outside the stellar surface.  This code solves the isothermal 1-D time-dependent fluid equations \citep{lyn74,pri81} in the thin disk approximation.  The output is the disk surface density, $\Sigma(r,t)$, as a function of radius, $r$, at selected times, $t=t_i$, as well as the mass flow rate as a function of position ${\dot M}(r)$.  This surface density is converted to volume density using the usual vertical hydrostatic equilibrium solution (a Gaussian) with a power law scale height, $H = h_0(r/R)^{1.5}$, where $R$ is the stellar equatorial radius.  The volume density then is used as the input to our 3-D Monte Carlo radiative transfer code, {\sc hdust} \citep{car06}.  {\sc hdust} simultaneously solves the NLTE statistical equilibrium rate equations and radiative equilibrium equation to obtain the hydrogen level populations, ionization fraction, and electron temperature as a function of position.  Its output is the emergent SED (and other observables of interest), which is used to calculate the $V$-band excess, $\Delta V$, of the disk as a function of time.  We assume that this variable disk emission dominates any intrinsic variability of the star.



We determine the $V$-band excess of 28\,CMa at the onset of the steep decline in late 2003 (time $t_0$) by calculating the average pre-decline brightness ($m_V=3.63\pm0.05$) and subtracting the average near the end of the quiescent phase ($m_V=4.18\pm0.02$).  This gives an initial excess $\Delta V(t_0) = -0.54 \pm 0.05$, where the error is estimated from the standard deviation of the means.  Although the disk building typically is not steady, the growth time required to produce this entire excess $t_\mathrm{g} = 0.54\,\mathrm{mag}/0.02\,\mathrm{mag\,d}^{-1}=0.07\,\mathrm{yr}$.  

Since the brightness of  28\,CMa  was relatively steady for two years prior to the decline, we assume the initial surface density of the disk was nearly the steady-state solution, $\Sigma(r,t_0) = \Sigma_0 (r/R)^{-2.0}$ \citep{bjo05}.  To match the initial $\Delta V$, we require $\Sigma_0 = 1.6 \pm 0.5 \,\mathrm{g}\,\mathrm{cm}^{-2}$, whose error is dominated by the uncertainty in the inclination angle ($i=0^\circ$--$30^\circ$).  This implies a pre-decline disk mass $M_\mathrm{d}=2\pi\Sigma_0 R^2\ln(r/R) = 3\times 10^{-9}\,M_\sun$ interior to $r=10\,R$.

\begin{figure}[!t]
\centerline{\includegraphics[width=0.8\linewidth]{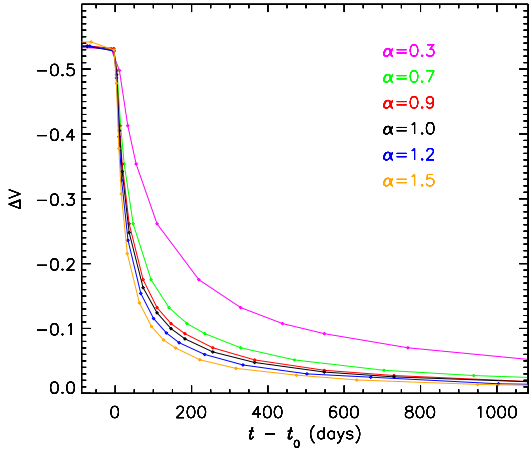}}
\caption{Disk dissipation.  Shown is the $V$-band excess as a function of time after mass injection stops for different values of $\alpha$ as indicated.  
}
\label{fig:DiskDissipation}
\end{figure}

To produce the initial steady-state disk, we run {\sc singlebe} for a long time with a constant mass injection rate, $\dot M$.  Note that $\Sigma_0$ depends on the ratio ${\dot M}/\alpha$, so whenever we change the disk viscosity parameter, $\alpha$, we adjust $\dot M$ to maintain the desired value for $\Sigma_0$.  After {\sc singlebe} achieves steady-state, we turn off the mass injection (by setting $\dot M = 0$) and let the disk dissipate.  
Figure~\ref{fig:DiskDissipation} shows the decline in the $V$-band excess as a function of time.  Note that as the viscosity parameter $\alpha$ increases, the disk dissipates more rapidly.  Initially there is a rapid decline as the inner disk (where the $V$-band excess is produced) adjusts from outflow to infall and is reaccreted.  Subsequently, the remaining mass of the disk is drained via quasi-steady-state accretion, and the $V$-band excess continues to slowly drop to zero.

\begin{deluxetable}{cccccc}
\tablecaption{Model Fits \label{tab:ModelFits}}
\tablewidth{0pt}
\tablehead{
\colhead{$\alpha$} &
\colhead{${\dot M}$} &
\colhead{$t_0$} &
\colhead{$\chi^2_\mathrm{I}$} &
\colhead{$\chi^2_\mathrm{II}$} &
\colhead{$\chi^2$} \\
\colhead{} &
\colhead{($M_\sun\ \mathrm{yr}^{-1}$)} &
\colhead{(MJD)} &
\colhead{} &
\colhead{} &
\colhead{} \\[-10pt]
}
\startdata
0.3 & $1.1 \times 10^{-8}$ & 52824 & 7.6 & 2.7  & 3.8 \\
0.8 & $2.8 \times 10^{-8}$ & 52929 & 1.4 & 2.1 & 2.0\\
0.9 & $3.2 \times 10^{-8}$ & 52933 & 1.2 & 2.3 & 2.1\\
1.0 & $3.5 \times 10^{-8}$ & 52935 & 1.2 & 2.4 & 2.2\\
1.2 & $4.3 \times 10^{-8}$ & 52939 & 1.6 & 2.6  & 2.5\\
1.5 & $5.3 \times 10^{-8}$ & 52942 & 2.9 & 2.9  & 3.0\\[-6pt]
\enddata
\end{deluxetable}

Given the model curves in Figure~\ref{fig:DiskDissipation}, we fit the Phase~I portion of the observed light curve (see Fig.~\ref{fig:LightCurve}) by adjusting only two parameters: $\alpha$ and $t_0$.  Table~\ref{tab:ModelFits} lists the best-fitting value of $t_0$ for each value of $\alpha$ and the reduced chi-squared value for the Phase~I data, $\chi^2_\mathrm{I}$. Additionally, we list the reduced chi-squared values of the extrapolation of the fit through Phase~II, $\chi^2_\mathrm{II}$, as well as the reduced chi-squared for the entire data set, $\chi^2$.  Finally, each model curve is shown in comparison to the observations in Figure~\ref{fig:LightCurve}.  The overall best fit to the Phase~I data is $\alpha = 1.0\pm0.2$ (90\% confidence interval).   
Note that when we extrapolate the fit to Phase~II, the fit is still reasonable,  requiring a large value of $\alpha$, albeit with a larger uncertainly.

\section{Discussion}
\begin{figure}[!t]
\centerline{\includegraphics[width=.8\linewidth]{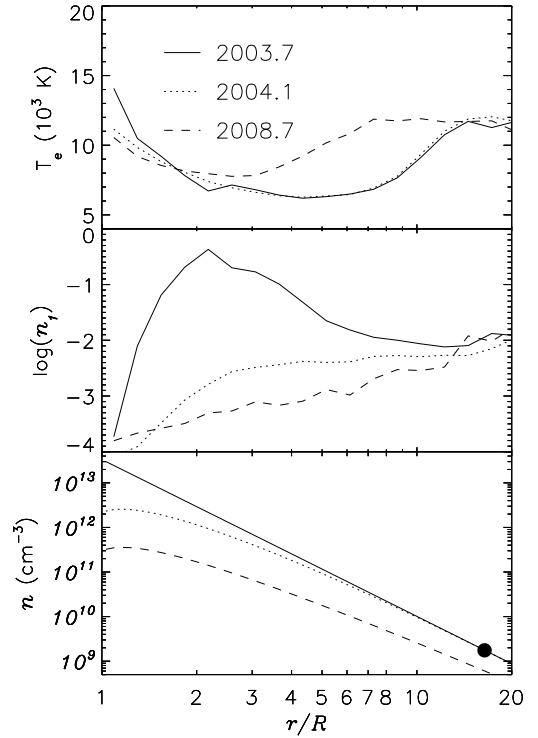}}
\caption{Disk Structure.  Shown are the disk temperature (top panel), hydrogen ground state population (middle panel), and density (bottom panel) as a function of radius for the initial state of the disk (solid line), at the end of the Phase~I (dotted line) and at the end of the simulation (dashed line) for the $\alpha=1.0$ case. The circle marks the position of the stagnation point (the division between inflow and outflow) for the 2004.1 curve.
}
\label{fig:DiskStructure}
\end{figure}

As may be seen in Figure~\ref{fig:LightCurve}, our model fits quite well the detailed shape of the $V$-band decline observed in 28\,CMa. Since the $V$-band excess is produced in the innermost parts of the disk \citep[typically $r \lesssim 2\,R$,][]{car11}, we infer that the viscous decretion disk model correctly predicts the time-dependent evolution of the disk density (at least at small radii).  Not only does the model reproduce the rapid early decline (Phase~I), it also reproduces the slow disk-draining (Phase~II) with a single physical model.  Furthermore, there is no evidence that the value of $\alpha$ changed from Phase~I to Phase~II,
despite the rather large change in physical conditions in the disk shown in Figure~\ref{fig:DiskStructure}.

When the mass injection ceases, the disk readjusts most quickly at small radii.  This is because the viscous diffusion timescale, $t_\mathrm{diff} \propto r^{1/2} \alpha^{-1} $, increases with radius.  Reaccretion of material from the inner disk occurs because the turbulent viscosity transports angular momentum outward, transferring angular momentum from the inner disk to the outer. Hence, the outer parts of the disk continue to expand while the inner parts move inward.  This produces simultaneous inflow in the inner disk and outflow in the outer disk.  As a function of time, the stagnation point (the division between inflow and outflow) moves outward, controlled by the diffusion time.  Interior to the stagnation point, the disk adopts a quasi-steady accretion solution, while exterior to the stagnation point the disk remains in quasi-steady decretion (see Fig.~\ref{fig:DiskStructure}).  Hence the outer disk acts as a mass reservoir that feeds reaccreting material through the inner disk where the $V$-band excess is produced.

We conclude that the decline rate in Phase~I is controlled by the viscous diffusion time; this is what determines the value of $\alpha$ in our model fit.  Since $\alpha$ is the only free physical parameter that we can vary, the fact that we can independently reproduce the rate of decline in Phase~II (when material is being transferred from the outer through the inner disk) is a non-trivial test of the time-dependent viscous decretion disk model.

In our analysis we assume, during the dissipation, that the disk is not being fed and no other important mass loss occurs.  Two such mass loss mechanisms are disk ablation and entrainment by the stellar wind, 
but general estimates of their mass loss rates \citep{krt11,gay99} indicate they are much smaller than that of the stellar wind. 
The disk reaccretion rate during Phase I is much larger than the wind mass loss rate, owing to the large value of $\alpha$, consequently neither plays an important role.

Perhaps the most surprising aspect of our results is the extremely large viscosity parameter ($\alpha \approx 1.0$) that is required.  In the $\alpha$-disk model \citep{sha73}, $\alpha$ parameterizes the turbulent (eddy) viscosity, $\nu=v_{\rm turb} \ell = \alpha c_s H$, where $v_{\rm turb}$ is the flow speed of the eddies and $\ell$ is their size scale.  Generally, one assumes that supersonic turbulence will not occur because of shock dissipation, so Shakura \& Sunyaev suppose the maximum flow speed is the sound speed, $c_s$, and that the characteristic size of the circulation cells is the disk pressure scale height, $H$.  Hence one expects $0 < \alpha < 1$.  To produce $\alpha = 1.0$ requires either 
roughly sonic flow and/or somewhat larger circulation cells.
If in fact $v_{\rm turb} \approx c_s$, this would strongly suggest that the origin of the turbulent mixing is an instability within the disk that grows until shock dissipation halts the growth.  Such an instability would also explain why $\alpha \approx 1$ regardless of the physical conditions within the disk.

Having now directly measured the value of $\alpha$ from the $V$-band decline rate, we can determine the mass decretion rate of 28\,CMa.  Normally one cannot use the disk density to determine the mass loss rate because the radial flow speeds are too small to be measured from the line profiles \citep{han00}.  Similarly, from a theoretical point of view, the disk surface density depends on the ratio ${\dot M} / \alpha$, so one must know $\alpha$ to find ${\dot M}$.  Assuming the pre-decline value of $\alpha$ is the same as that we measure during disk dissipation ($\alpha = 1.0$), we find that the mass injection rate of 28\,CMa was ${\dot M} = (3.5 \pm 1.3) \times 10^{-8}\ M_\sun\ \mathrm{yr}^{-1}$.  An important check of this rate can be made using the growth time, $t_\mathrm{g}$, in combination with the pre-decline disk mass, $M_\mathrm{d}$, which implies ${\dot M} \sim M_\mathrm{d}/t_\mathrm{g} = 4 \times 10^{-8}\ M_\sun\ \mathrm{yr}^{-1}$ during the growth phase.  Consequently, we infer there is no evidence for a significant change in $\alpha$ between the disk growth and the disk dissipation phases.

We find, therefore, that the mass injection rate is much larger than previously thought, and is at least an order of magnitude larger than the mass loss in the stellar wind, as inferred from UV observations \citep{sno81}.  This implies that the stellar wind is not the mechanism responsible for mass injection into the disk, unless the wind can be prevented from producing strong, easily observable UV wind lines.

\section{Conclusions}

Light curve fitting to the dissipation of a disk can be a powerful tool for measuring $\alpha$. Further it provides additional quantitative tests of the viscous decretion disk model.  Here we have shown that the viscous decretion disk model can reproduce the detailed time-dependence of the dissipation of a Be star disk.  Using the $V$-band excess to measure how the disk density decreases with time, we have been able for the first time to directly measure the value of the disk viscosity parameter.  We find $\alpha = 1.0\pm0.2$, which provides an important clue about the origin of the turbulent viscosity, suggesting that it likely is produced by an instability in the disk whose growth is limited by shock dissipation.

  Finally, knowing the value of $\alpha$, we 
   calculate that the mass decretion rate
   of 28\,CMa was ${\dot M} = (3.5\pm1.3) \times 10^{-8}\ M_\sun\ \mathrm{yr}^{-1}$.  Such a large required mass injection rate places strong constraints on any proposed stellar mass loss mechanism.

\acknowledgments
We thank the referee, Stanley P. Owocki, for his useful comments. We acknowledge support from CNPq grant 308985/2009-5 (ACC) and Fapesp grants 2010/19029-0 (ACC), 2010/16037-2 (JEB), and 2009/07477-1 (XH).

%


\begin{thebibliography}{}

\bibitem[Ando(1986)]{and86} Ando, H. 1986, A\&A, 163, 97




\bibitem[Argelander(1843)]{arg1843} Argelander, F.W.A.\ 1843, Uranometria Nova, Berlin

\bibitem[Baade et al.(2011)]{baa01} Baade, D., Rivinius, T., 
{\v S}tefl, S., \& Martayan, C.\ 2011, IAU Symposium, 272, 1 

\bibitem[Bjorkman \& Carciofi(2005)]{bjo05} Bjorkman, J.E. \& Carciofi, A.C. 2005, in ASP Conf Ser. 337, The Nature and Evolution of Disks Around Hot Stars, ed R. Ignace \& K. Gayley (San Francisco: ASP), 75

\bibitem[Carciofi \& Bjorkman(2006)]{car06} Carciofi, A.~C. \& Bjorkman, J.~E.\ 2006, \apj, 639, 1081


\bibitem[Carciofi et al.(2009)]{car09} Carciofi, A.~C., Okazaki, A.~T., Le Bouquin, J.-B., et al.\ 2009, \aap, 504, 915 

\bibitem[Carciofi (2011)]{car11} Carciofi, A.~C.\ 2011, IAU Symposium, 272, 325 

 
\bibitem[Cranmer(2005)]{cra05} Cranmer, S.~R.\ 2005, \apj, 
634, 585 

\bibitem[Cranmer(2009)]{cra09} Cranmer, S.~R.\ 2009, \apj, 701, 396

\bibitem[Doazan et al.(1983)]{doa83} Doazan, V., Franco, M., Rusconi, L., Sedmak, G., \& Stalio, R. 1983, A\&A, 128, 171

\bibitem[Gayley et al.(1999)]{gay99} Gayley, K.~G., Owocki, 
S.~P., \& Cranmer, S.~R.\ 1999, \apj, 513, 442

\bibitem[Hanuschik(2000)]{han00} Hanuschik, R.~W.\ 2000, IAU 
Colloq.~175: The Be Phenomenon in Early-Type Stars, 214, 518 


\bibitem[Kraus et al.(2011)]{kra11} Kraus, S., Monnier, 
J.~D., Che, X., et al.\ 2011, arXiv:1109.3447 

\bibitem[Krti{\v c}ka et 
al.(2011)]{krt11} Krti{\v c}ka, J., Owocki, S.~P., \& Meynet, G.\ 2011, \aap, 527, A84 

\bibitem[Lee et al.(1991)Lee, Osaki, \& Saio]{lee91} Lee, U., Osaki, Y., \& Saio, H. 1991, \mnras, 250, 432

\bibitem[Lynden-Bell \& Pringle(1974)]{lyn74} Lynden-Bell,~D., \& Pringle,~J.~E. 1974, \mnras, 168, 603

\bibitem[Maintz et al.(2003)]{mai03}Maintz,~M. Rivinius,~Th., \v{S}tefl,~S., Baade,~D., Wolf,~B., \& Townsend,~R.~H.~D. 2003, A\&A, 411, 181

\bibitem[Neiner et al.(2002)]{nei02} Neiner, C. 2002, A\&A, 388, 899

\bibitem[Oudmaijer et al.(2011)]{oud11} Oudmaijer, R.~D., 
Wheelwright, H.~E., Carciofi, A.~C., Bjorkman, J.~E., 
\& Bjorkman, K.~S.\ 2011, IAU Symposium, 272, 418 


\bibitem[Okazaki et al.(2002)]{oka02} Okazaki, A.~T., Bate, M.~R., Ogilvie, G.~I., \& Pringle, J.~E.\ 2002, \mnras, 337, 967 

\bibitem[Okazaki(2007)]{oka07} Okazaki, A.~T.\ 2007, Active OB-Stars: Laboratories for Stellar and Circumstellar Physics, 361, 230 

\bibitem[Otero et al.(2001)Otero, Fraser, \& Lloyd]{ote01} Otero,~S.~A., Fraser,~B., \& Lloyd,~C. 2001, Information Bulletin on Variable Stars, 5026, 1 

\bibitem[Otero \& Moon(2006)]{ote06} Otero,~S.~A., \& Moon,~T. 2006, The Journal of the American Association of Variable Star Observers, 34, 156 

\bibitem[Owocki(2006)]{owo06} Owocki, S.\ 2006, Stars with 
the B[e] Phenomenon, 355, 219 


\bibitem[Porter \& Rivinius(2003)]{por03} Porter, J.M. \& Rivinius, T. 2003, PASP, 115, 1153


\bibitem[Pringle(1981)]{pri81} Pringle,~J.~E. 1981, ARA\&A, 19, 137

\bibitem[Rivinius et al.(1998)]{riv98} Rivinius, Th., Baade, D., Stefl, S., Stahl, O., Wolf, B., \& Kaufer, A. 1998, A\&A, 333, 125

\bibitem[Shakura \& Sunyaev(1973)]{sha73}
  Shakura N.~I., Sunyaev R.~A., 1973, \aap, 24, 337
  
\bibitem[Snow(1981)]{sno81} Snow, T.~P., Jr.\ 1981, \apj, 
251, 139 

\bibitem[{\v S}tefl et al.(2003)]{ste03} {\v S}tefl, S., Baade, D., Rivinius, T., et al.\ 2003, \aap, 402, 253 

\bibitem[{\v S}tefl et al.(2009)]{ste09} {\v S}tefl, S., Rivinius, T., Carciofi, A.~C., et al.\ 2009, \aap, 504, 929 

\bibitem[Underhill \& Doazan(1982)]{und82} Underhill, A. \& Doazan, V. eds. 1982, B Stars with and without Emission Lines (NASA SP-456; Washington, DC: NASA)

\bibitem[Wisniewski et al.(2010)]{wis10} Wisniewski, J.~P., 
Draper, Z.~H., Bjorkman, K.~S., et al.\ 2010, \apj, 709, 1306

\end{thebibliography}
\end{document}